\documentclass[aps,prl,twocolumn,amsmath,amssymb]{revtex4-1}
\usepackage{graphicx}
\usepackage{subfigure}
\usepackage{epsfig}
\usepackage{dcolumn}
\usepackage{bm}
\usepackage{color}
\usepackage{amsbsy}
\usepackage{hyperref}
\usepackage{booktabs}

\def\be{\begin{equation}}       \def\ee{\end{equation}}
\def\bea{\begin{eqnarray}}      \def\eea{\end{eqnarray}}
\def\ba{\begin{array}}
\def\ea{\end{array}}
\def\bnum{\begin{enumerate} }
\def\enum{\end{enumerate}}

\def\=>{\Rightarrow}
\def\>{\rightarrow}

\def\eye2{Fathbb{I}}

\renewcommand{\v}[1]{{\bf #1}}

\newcommand{\si}{{\sigma}}

\renewcommand{\>}{\rangle}

\newcommand{\al}{\alpha}
\newcommand{\RNum}[1]{\uppercase\expandafter{\romannumeral #1\relax}} 

\begin{document}

\title{$n$-Hourglass Weyl fermions in nonsymmorphic materials}
\author{Yijie Zeng}
\author{Luyang Wang}
\email{wangluyang730@gmail.com}
\author{Dao-Xin Yao}
\email{yaodaox@mail.sysu.edu.cn}
\affiliation{State Key Laboratory of Optoelectronic Materials and Technologies, School of Physics, Sun Yat-Sen University, Guangzhou 510275, China}

\begin{abstract}
Hourglass-like band structures protected by nonsymmorphic space group symmetries can appear along high-symmetry lines or in high-symmetry surfaces in the Brillouin zone. In this work, from symmetry analysis, we demonstrate that $n$-hourglass-like band structures, a generalization of hourglass-like band structures, which host a number of Weyl points, are enforced along screw-invariant lines in non-magnetic materials with a single $N$-fold screw axis when spin-orbit coupling is finite, where $n$, a non-unity factor of $N$, denotes the degree of the screw-invariant line. The ``standard" $n$-hourglass has $n-1$ crossings, which is minimal, and its variants can have more crossings. Purely by symmetry considerations, we find there are minimally two particle and two hole Fermi pockets enclosing Weyl points with opposite monopole charges at proper fillings, which can result in distinct physical effects including the possible formation of topological density waves and the quantum nonlinear Hall effect. We construct a minimal model which respects all the symmetries, and from which we see how the $n$-hourglasses appear when spin-orbit coupling is turned on. The same results are derived from compatibility relations. As exemplary, BiPd, ZnTe under high pressure, and the high-temperature phase of Tl$_3$PbBr$_5$, are shown from first-principles calculations to exhibit $n$-hourglass-like band structures, with $N=2,3,4$, respectively, which confirms our symmetry analysis and minimal model.
\end{abstract}
	
\date{\today}
\maketitle

{\it Introduction}.---The interplay between symmetry and topology has been a hot topic in condensed matter physics in recent years, stimulated by the discovery of topological insulators\cite{Hasan2010,Qi2011}. Topological insulators have been classified in the tenfold way according to whether they respect time-reversal symmetry (TRS), particle-hole symmetry and chiral symmetry\cite{Schnyder2008,Kitaev2009}. On the other hand, the theory of symmetry-protected topological (SPT) phases has been developed from the perspective of quantum information\cite{Chen2013}, which feature short-range quantum entanglement and can be smoothly deformed to the trivial gapped phase if the symmetry is allowed to be broken. From this perspective, topological insulators are the most studied SPT phase.

The discovery of topological crystalline insulators\cite{Fu2011} has led to the realization that crystalline symmetries can vastly enrich topological phases. For crystalline solids with topological degeneracies, by digging into the 230 space groups (SGs), fermions beyond the Weyl-Dirac-Majorana classification have been found\cite{Bradlyn2016}, and the filling constraints of semimetals in each SG have been presented\cite{Watanabe2016}. When interaction is considered, novel phenomena may occur, such as the crystalline symmetry fractionalization\cite{Qi2015PRL}.

Nonsymmorphic SG symmetries play a special role in SPT phases. For gapped systems, nonsymmorphic topological insulators and superconductors are classified according to their nonsymmorphic symmetries\cite{Shiozaki2016}. In particular, the surface states of nonsymmorphic topological insulators show hourglass-like dispersions\cite{Wang2016Nature}. The hourglass-like dispersions also appear in the bulk Brillouin zone (BZ) of nonsymmorphic materials with glide planes or twofold screw axes\cite{Young2015}, and may result in nodal points\cite{Wang2017PRB,Furusaki2017}, nodal lines\cite{Chen2016,Takahashi2017,Wang2017PRB}, nodal chains\cite{Bzdusek2016,Wang2017NC}, nodal nets\cite{Singh2018PRL,Fu2018PRB} and nodal surfaces\cite{Wu2018}. However, the topological degeneracies in systems which have a screw axis with degree more than two have not been well explored yet.

In this work, we investigate the spin-orbit coupled crystalline solids which have a single $N$-fold screw axis and respects TRS. By symmetry analysis, we find that the symmetries enforce $n$-hourglass-like dispersion along screw-invariant lines, which is a generalization of the hourglass-like dispersion. In this language, the conventional hourglass is a 2-hourglass, and 3-, 4- and 6-hourglass can appear in systems with nonsymmorphic SG symmetries. The integer $n>1$ is a factor of $N$. The SGs under consideration are $P2_1$, $P3_1$, $P3_2$, $P4_1$, $P4_2$ and $P4_3$, all of which have a single screw axis and are non-magnetic. We notice that the SGs $P6_i$ with $i=1,2,...,5$ (No. 169-173) have been considered in Ref.\cite{Zhang2018}. A minimal model is presented to produce the $n$-hourglass-like dispersion for an arbitrary $n$.

The crossings of the $n$-hourglass are Weyl points\cite{Armitage2018}, while the Kramers degeneracies at its edges can either be Weyl points or reside on a nodal plane, depending on the symmetries. We focus on the crossings, and find that there can be a minimal number of two Weyl points in half of the Brillouin zone for $N=3$ and 4. The possible physical effects, such as the topological density waves and the nonlinear quantum Hall effect, are discussed.

We perform first-principles calculations of band structures on three exemplary materials, BiPd (SG $P2_1$), ZnTe under high pressure (SG $P3_1$) and Tl$_3$PbBr$_5$ at high temperature (SG $P4_1$), which show $n$-hourglass-like band structures for $N=2,3,4$, respectively. Their band structures and the location of the Weyl points are completely consistent with our symmetry analysis.

{\it Symmetry analysis}.---We study non-magnetic crystalline systems with a single $N$-fold screw axis with a sizable spin-orbit coupling (SOC). The screw operations can be ${\widetilde C_{2z}}$, ${\widetilde C_{3z}^{1,2}}$ and ${\widetilde C_{4z}^{1,2,3}}$, corresponding to SGs $P2_1$ (No. 4), $P3_1$ (No. 144) and $P3_2$ (No. 145), and $P4_1$ (No. 76), $P4_2$ (No. 77) and $P4_3$ (No. 78), respectively.

\begin{figure}[t]
    \includegraphics[width=8cm]{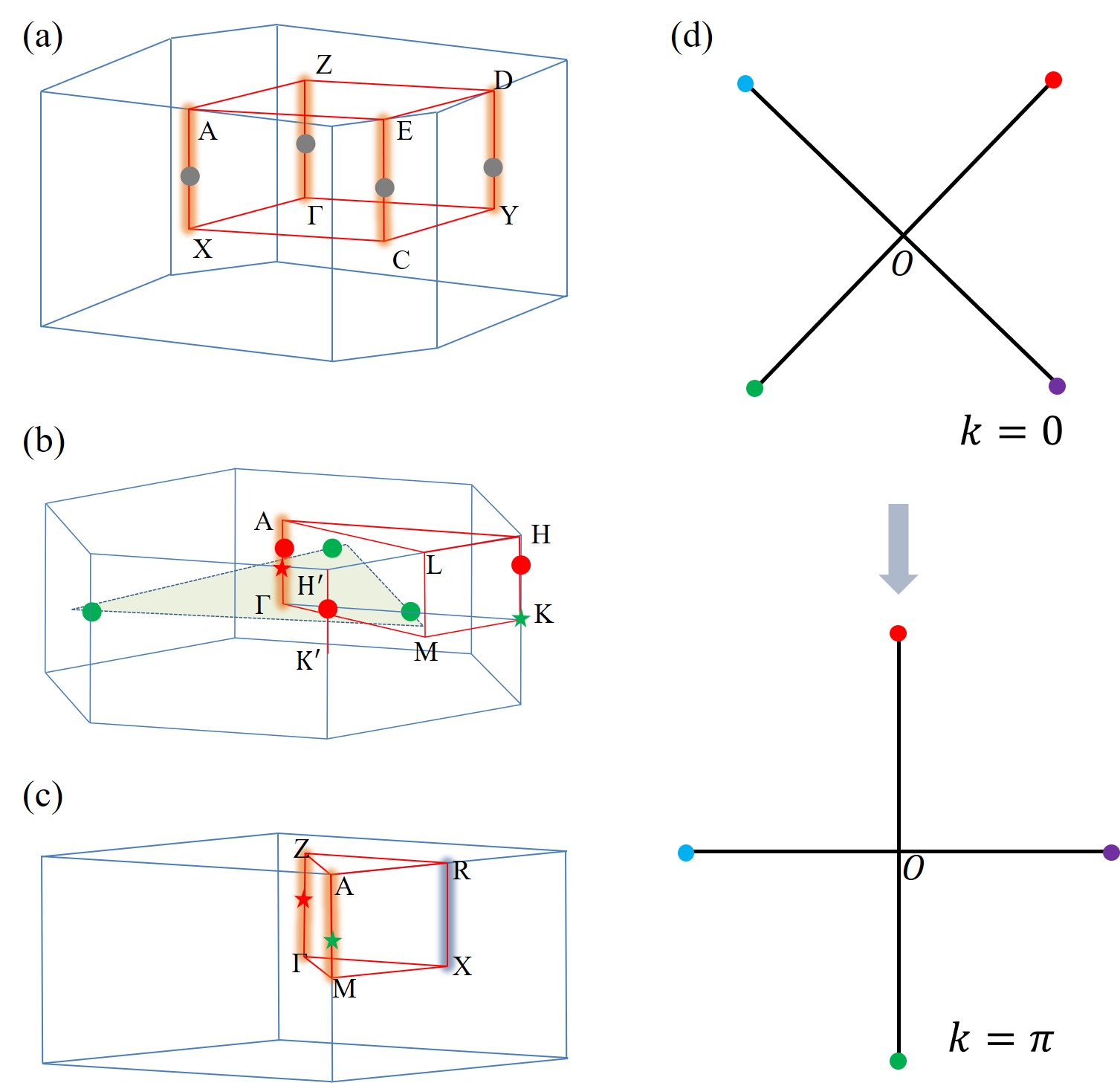}\label{fig1ad}\vspace{0.5cm}
    \includegraphics[width=8cm]{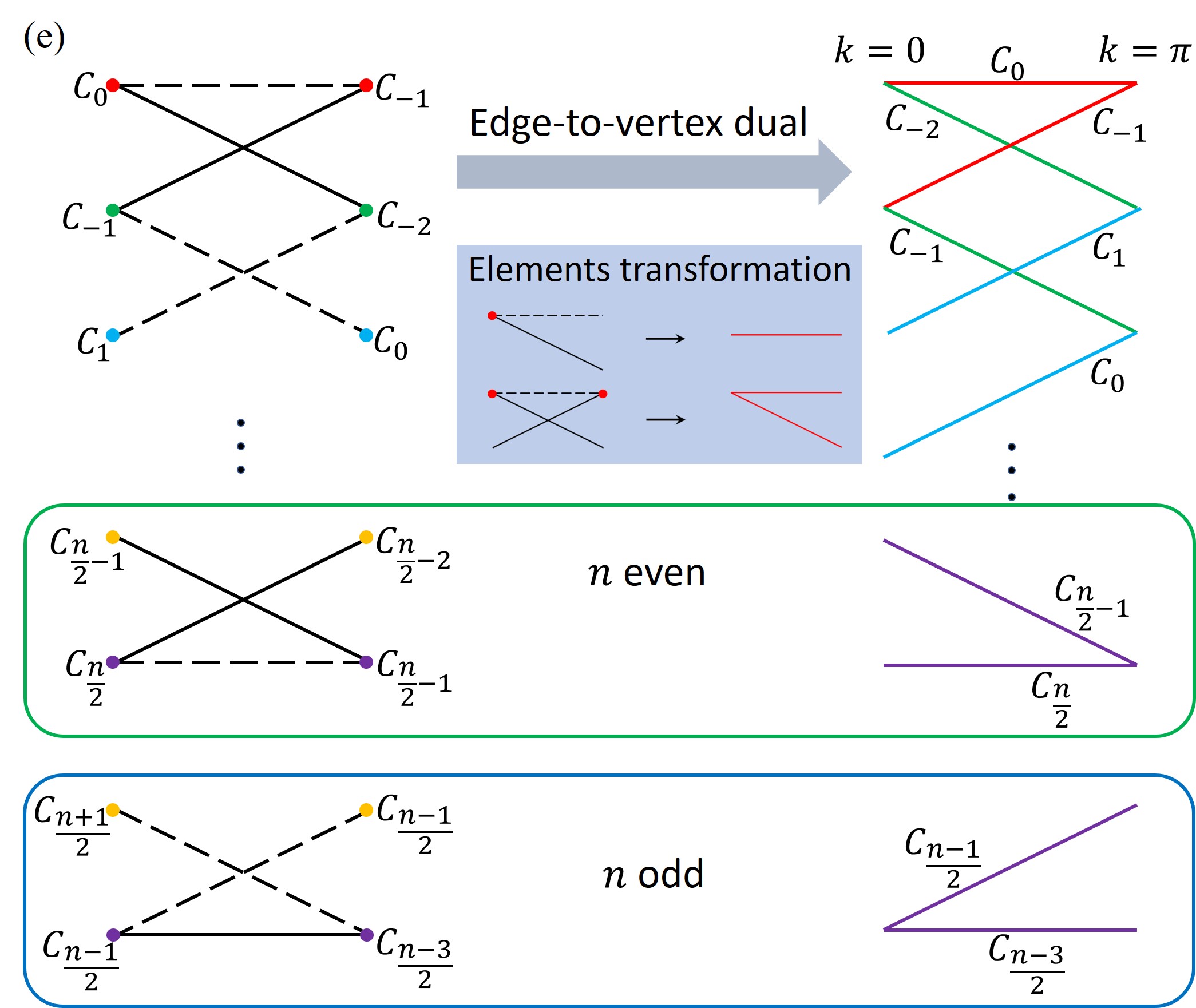}\label{fig1e}
	\caption{(a-c) The BZ of SG $P2_1$ (a), $P3_1$ (b) and $P4_1$ (c), which belong to the simple monoclinic, simple hexagonal, and simple tetragonal Bravais lattice, respectively. (d) Eigenvalues of the screw rotation for $n=4$ shown in the complex plane at $k=0$ and $\pi$. The rotation of an angle $\pi/n$ during the path going from $k=0$ to $k=\pi$ is the origin of the $n$-hourglass. (e) Schematic of how the $n$-hourglass results from different patterns of Kramers pairing at $k=0$ and $\pi$, for both $n$ even and $n$ odd. $C_{m}$ means the $m$ (mod $n$)-th eigenvalue of $\widetilde{C}_{nz}^1$, $C_{n,m}$.}
	\label{fig:graph}
\end{figure}

TRS in spinful systems imposes Kramers degeneracy at time-reversal invariant momenta (TRIM). In symmorphic materials, the band connections between two TRIM are trivial. As a result, symmorphic insulators have fillings $2\nu$ where $\nu$ is an integer. However, in nonsymmorphic systems, glide or screw symmetries lead to nontrivial band connections. For instance, glide and twofold screw symmetries can give rise to hourglass-like dispersions\cite{Wang2017PRB}, due to the different pairing patterns of the symmetry eigenvalues at different TRIM.

In a system with both TRS and an $N$-fold screw axis where $N>2$, band structures similar to the hourglass, but with more bands involved in a set, generically appear. This is the foundation of the filling-enforced gaplessness\cite{Watanabe2016}. Here we explore the topological degeneracies in detail from symmetry considerations.

The high symmetry points and axes in the first BZ are shown in Fig.\ref{fig:graph}(a-c) for $N=2,3,4$, respectively. Two TRIM along the direction of the screw axes may or may not be connected by a screw-invariant line. The latter case appears in the SGs with $N=3$: M and L are TRIM, but ML is not screw-invariant, so the band connection between M and L is trivial. On the other hand, KH is screw-invariant, but K and H are not TRIM, so crossings seem not guaranteed (but actually indirectly enforced by symmetries, as shown later). We mainly focus on the cases where two TRIM are connected by screw-invariant lines, which are indicated by the shaded lines in Fig.\ref{fig:graph}(a-c). There are four such lines for $N=2$, one for $N=3$ and three for $N=4$. Note that for $N=4$, there are two fourfold screw-invariant lines and one twofold screw-invariant line, shown by shadings of different colors. Therefore, when we discuss $n$-fold screw-invariant lines in the {\it momentum} space in systems with an $N$-fold screw axis in the {\it real} space, we are aware of that $n$ is a non-unity factor of $N$.

We focus on the screw rotations $\widetilde C_{nz}^1$ and use $C_{n,m}$ to label its eigenvalues, while other types of screw, such as that in SG $P4_2$, will be analyzed in a similar way later. Since $n$ times of screw operations translate the system by a unit cell along the screw axis, we have $(\widetilde C_{nz}^1)^n=-e^{ik}$ where the lattice constant along the $z$-direction is assumed to be 1, $k$ is the wave vector in that direction, and the minus sign accounts for the rotation of spin-1/2. Therefore, the eigenvalues of $\widetilde C_{nz}^1$ are $C_{n,m}(k) =e^{\frac{i(k+\pi)+i2\pi m}{n}}$, with $m=0,1,...,n-1$. Time-reversal operation pairs the states with eigenvalues conjugate to each other at the TRIM. At $k=0$, the states with $C_{n,m}(0)$ pair with those with $C_{n,n-m-1}(0)$; while at $k=\pi$, the states with $C_{n,m}(\pi)$ pair with those with $C_{n,n-m-2}(\pi)$. This is illustrated in the left panel of Fig.\ref{fig:graph}(e), where the dashed lines indicate the 0-pairing while the solid lines $\pi$-pairing, where $k$-pairing means the Kramers pairing at $k$. In graph theory, the lines are called edges and the points are called vertices. The different colors of the vertices indicate different eigenvalues of $\widetilde{C}_{nz}^1$, and the two possibilities how the graph ends are shown, depending on whether $n$ is even or odd. The sum of $m$ of the two eigenvalues is $-1$(mod $n$) if the vertices are connected by dashed lines, and $-2$(mod $n$) by solid lines.  Such pairing relations result in the dispersion shown in the right panel of Fig.\ref{fig:graph}(e) in the following way. Each vertex is transformed into an edge and each edge is transformed into a vertex. The colors are kept the same in the transformation for clearness. This is like the edge-to-vertex dual (or the line graph) in graph theory, but with the constraints that the vertices of the dual graph must be equally distributed at $k=0$ and $\pi$, and the crossings of the original graph are not considered as vertices. We call the dual graph the $n$-hourglass-like dispersion, where $n$ denotes the degree of the screw-invariant line, and $2n$ bands stick together. The eigenvalues at $k=0$ and $\pi$ in the complex plane are explicitly shown in Fig.\ref{fig:graph}(d) for $n=4$, the different patterns of pairing of which at the two TRIM lead to the $n$-hourglass. The same results can also be derived from the compatibility relations between the irreducible representations of the little group along the screw-invariant lines ended with TRIM, see Ref.\cite{SM} for details.

In graph-theoretical language, the $n$-hourglass is a (2,2)-biregular graph. The vertices of both graphs are divided into two subsets with an equal cardinality $n$. In the simplest case, the dual graph has $n-1$ crossings, which we call the ``standard" $n$-hourglass. All crossings are Weyl points in three-dimensional systems. Along the screw-invariant lines, we call the TRIM near the center of the BZ the inner edge of the $n$-hourglass, whereas the TRIM farther the outer edge. The $n$ vertices corresponding to the inner edge of the $n$-hourglass are also Weyl points. However, the outer edge may not host Weyl points, depending on the value of $n$. For $n=2$, it has been shown that the degeneracies at the outer edge reside on a nodal surface. This also happens for $n=4$. However, for $n=3$, the degeneracies at the outer edge are Weyl points. Permutations of the vertices at the edges will introduce more crossings, and curving the bands could also increase the number of Weyl points.

{\it Physical effects}.---The conventional hourglass Weyl fermions, corresponding to 2-hourglass in the language used here, has been studied in Ref.\cite{Wang2017PRB}, and here we show the distinct properties for $N=3$ and $4$. For $N=3$, there is only one screw-invariant line $\Gamma$A connecting two TRIM. Therefore, there is minimally one symmetry-enforced Weyl point along $\Gamma$A as the crossing between, for example, the second and the third band in a 3-hourglass. (Only the upper half of the BZ is considered, while the other half is related by the TRS.) Due to the fermion doubling theorem\cite{Nielsen1981}, there must exist an odd number of other Weyl points for the cancellation of the total monopole charge.  The threefold rotational symmetry tells us that Weyl points could triply appear with the same charge if located inside the BZ, which cannot make the sum of the monopole charges vanish by themselves. If the Weyl points are located at the surface of the BZ, they contribute $\pm3/2$ monopole charge. Therefore, the only possibility is that the other(s) is at another screw-invariant axis, i.e. KH or $\mbox{K}'\mbox{H}'$, or both. Then there are two possible cases: (i) One other Weyl point is located at KH {\it or} $\mbox{K}'\mbox{H}'$, which we call the 1+1 case; (ii) One other Weyl point is at KH {\it and} one at $\mbox{K}'\mbox{H}'$, then there must be a triple set of Weyl points inside the BZ, which we call the 3+3 case. In the 3+3 case, the triple set inside the BZ is indirectly enforced by symmetries.

The 1+1 case can also occur in SG $P4_1$. The symmetry-enforced crossings between the fourth and the fifth band in a 4-hourglass are located at the two 4-hourglasses along $\Gamma$Z and MA, but not along XR.

\begin{figure}[t]
	\includegraphics[width=8cm]{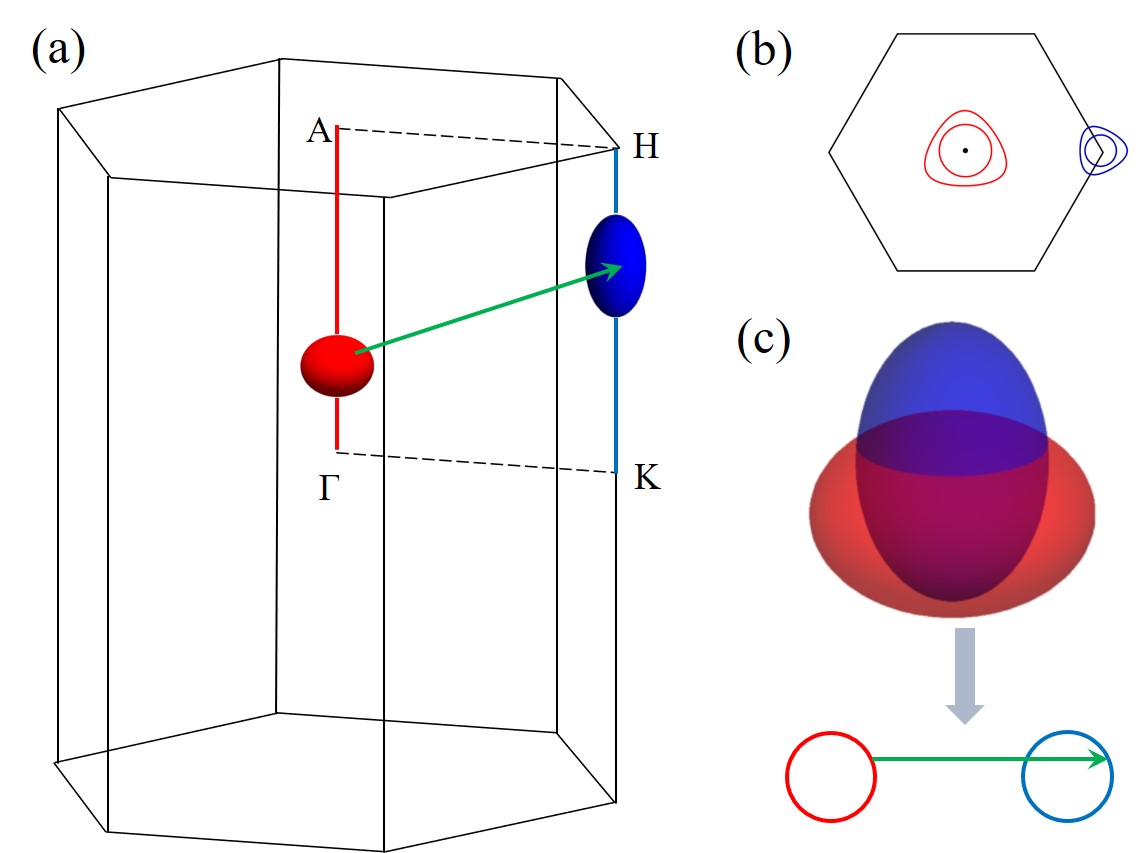}
	\caption{(a) Two Fermi pockets, one electron-like and one hole-like, at the two screw-invariant lines. Each pocket encloses a Weyl point. (b) The projection of the two Fermi pockets on the plane perpendicular to the $k_z$-axis at two different energies. The Fermi pockets have threefold rotational symmetry, which become circular if the Fermi energy is close to the Weyl point. (c) Schematic of how the nesting of Fermi surface submanifolds occurs. The green arrow in (a) indicate the nesting vector, whose projection is shown in (c), together with the projection of the Fermi pockets.}
	\label{fig:nesting}
\end{figure}

Now we focus on the 1+1 case for $N=3$ and $4$. Since the two Weyl points are not related by symmetries, they are not at the same energy\cite{Ruan2016NC,Ruan2016PRL,Wang2016PRA}. If the Fermi energy is tuned to be between them, e.g. $1/3$ filling for $N=3$ and $1/2$ filling for $N=4$, there will be a particle and a hole Fermi pocket, each enclosing a Weyl point with a monopole charge opposite to the other. A possible phenomenon is the formation of topological density waves\cite{WangYX2016}. The cross section of the Fermi pockets perpendicular to the $k_z$-axis has a threefold rotational symmetry. But if the Fermi pocket is close to the Weyl point, the trigonal warping vanishes and the cross section is a circle (like in graphene, see e.g. Ref.\cite{Neto2009}), resulting in an ellipsoidal Fermi pocket, as shown in Fig.\ref{fig:nesting}(a,b). Although there is not nesting between the two ellipsoids due to the lack of particle-hole symmetry, there must be nesting between submanifolds of the Fermi surfaces. As shown in Fig.\ref{fig:nesting}(c), the cross sections of the two Fermi pockets can be of the same size, resulting in Fermi surface nesting of two-dimensional (2D) character. When a repulsive interaction $V$ is taken into account, the nesting causes transition to charge or spin density waves at $T_c=\Lambda_c e^{-2/(N(0)V)}$ where $\Lambda_c$ is the high energy cutoff and $N(0)$ is the 2D density of states at the cross section; while if the interaction is an attractive $U$, pair-density wave instability occurs at $T_c=\Lambda_c e^{-2/(N(0)U)}$\cite{WangYX2016}. Intuitively, when the cross section is at the equator of the blue ellipsoid, we have the largest $N(0)$ and the system is the most susceptible, from which the nesting vector (the green arrows in Fig.\ref{fig:nesting}) is found. Although in general the ellipsoids are deformed since the Weyl cone is tilted in the $z$-direction, the cross sections are still circular, and the above argument always works. Detailed calculations are beyond the scope of the present work.

The materials in the SGs considered here preserve TRS and break inversion symmetry, thus cannot have the linear Hall effect. However, the tilted Weyl cones generate Berry curvature dipoles, which can lead to the quantum nonlinear Hall effect\cite{Sodemann2015,Ma2018}. This effect should be readily observed in the materials with $n$-hourglass Weyl fermions.

The crystals with SG $Pn_{n-1}$ symmetry have the same $n$-hourglass-like dispersions as $Pn_1$. So the only case not discussed yet is SG $P4_2$. It has not 4-hourglass-like dispersion, but only 2-hourglasses, similar to SG $P2_1$. This is consistent with the filling constraint for this SG\cite{Watanabe2016}.

\begin{figure}[t]
	\includegraphics[width=8cm]{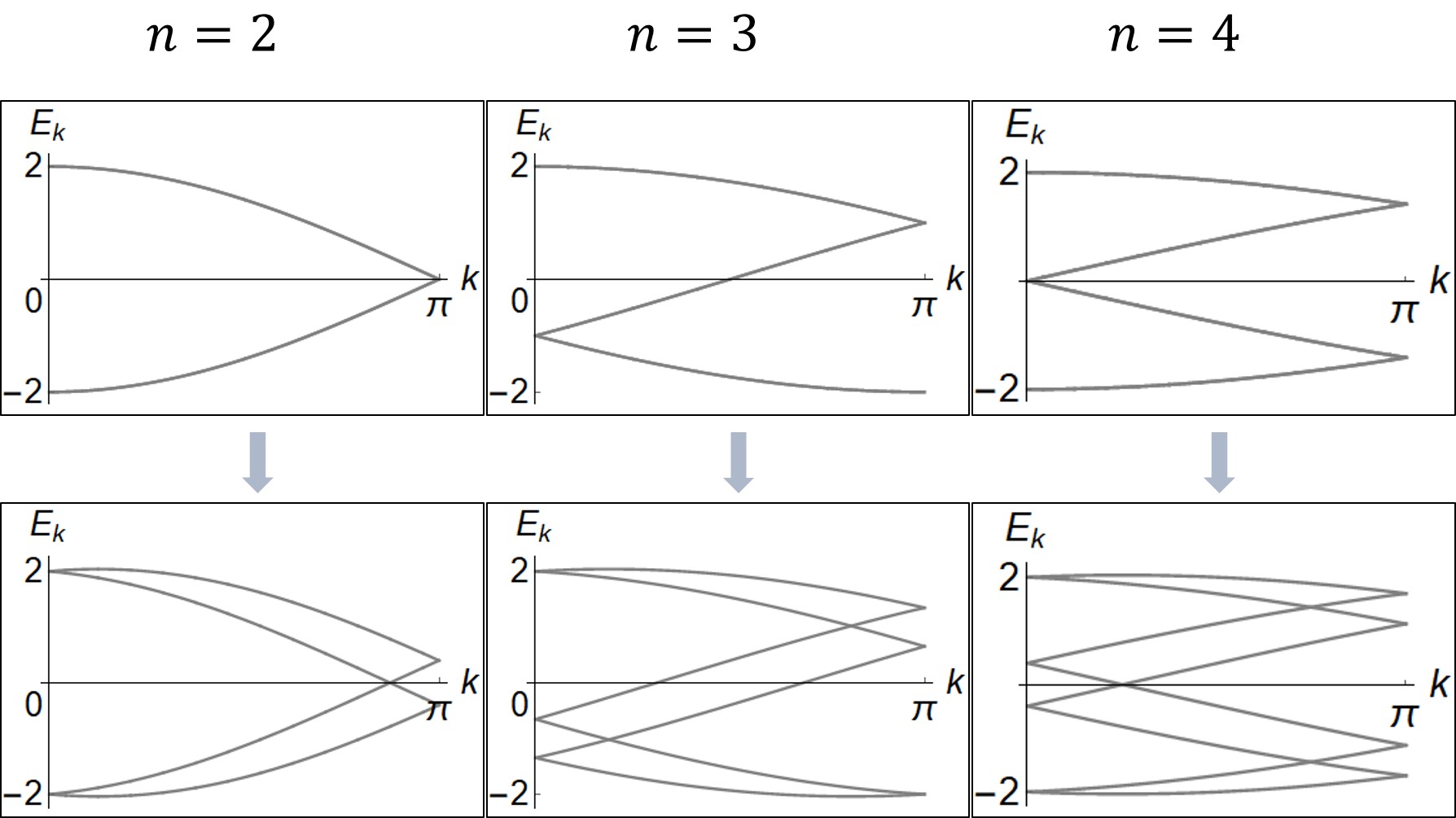}
	\caption{Schematic of how $n$-hourglasses emerge for $n=2,3,4$ when the SOC is turned on. $\al=0$ in the upper plots and $\al=0.2$ in the lower plots.}
	\label{fig:model}
\end{figure}

{\it Minimal model for the $n$-hourglass-like dispersion}.---The tight-binding model for $2$-hourglass-like dispersion has been constructed in Ref.\cite{Wang2017PRB}. Here, instead, we develop a minimal model for the $n$-hourglass-like dispersion for an arbitrary $n$ purely from symmetry considerations. The model reads
\begin{eqnarray}
  H_n(k) &=& S_{n}(k)+S_{n}^\dagger(k)+\al i(S_{n}(k)-S_{n}^\dagger(k))\si_x,
\end{eqnarray}
where $\si_i$ denotes the Pauli matrices acting on spin, $\al$ is a parameter characterizing the strength of SOC, and $S_n$ is an $n$-dimensional matrix representing the $n$-fold screw rotation,
\begin{eqnarray}
  S_n &=& \left(\begin{array}{cccc}0&0&\dots&e^{ik}\\1&0&\dots&0\\0&\ddots&0&0\\ 0&\dots&1&0\end{array}\right),
\end{eqnarray}
which represents a permutation of the $n$ atoms inside a unit cell followed by a translation of one atom along the screw axis by the lattice constant, an element in SG $Pn_1$. $H_n(k)$ constructed in this way satisfies the following conditions: (i) it is hermitian; (ii) it respects time-reversal and $n$-fold screw symmetry; and (iii) it has SOC included. In Fig.\ref{fig:model}, we show how the $n$-hourglasses emerge from the V-like, N-like and W-like dispersions, for $n=2,3,4$, respectively, upon the tuning of the SOC from zero to finiteness.

\begin{figure}[t]
	\includegraphics[width=8cm]{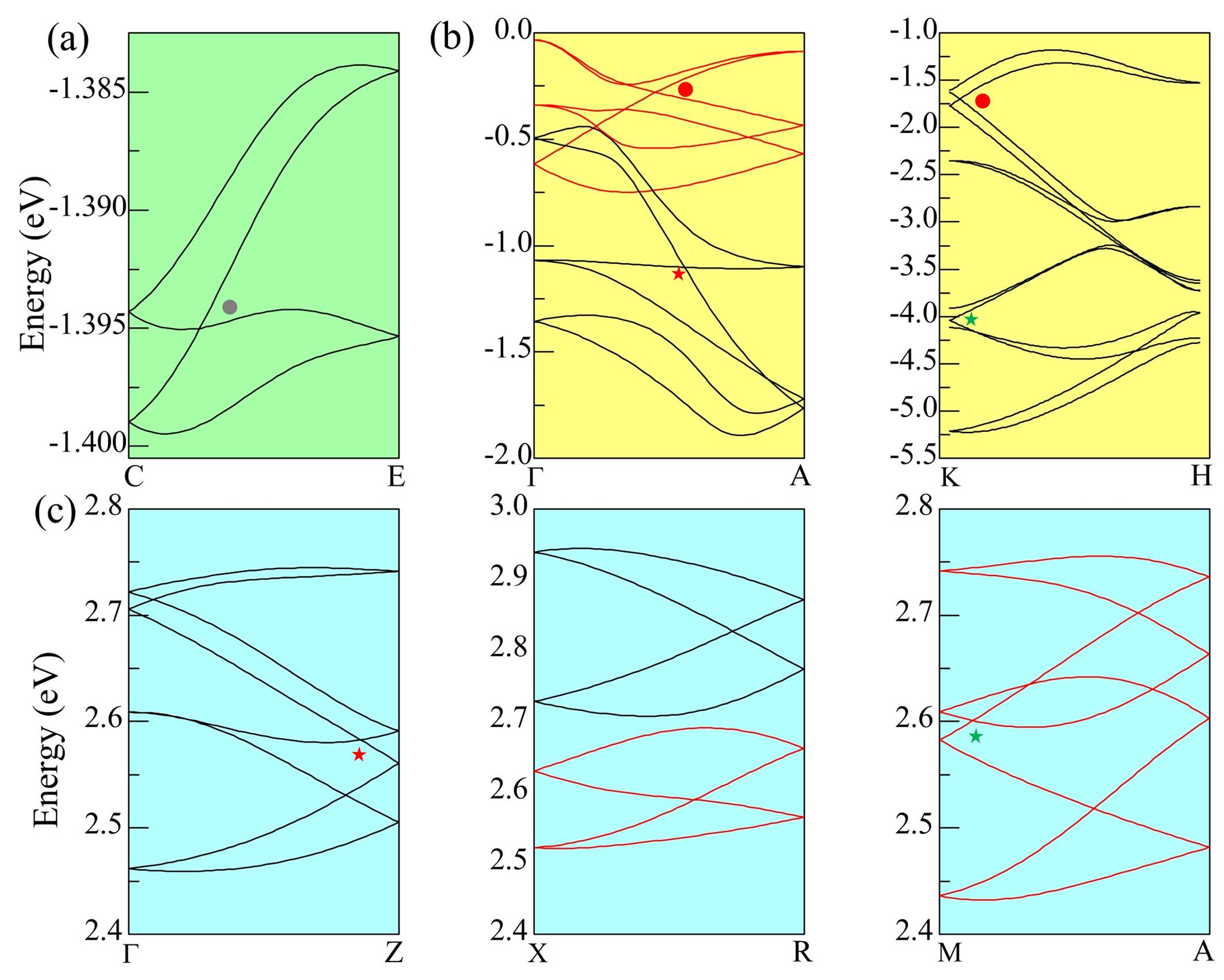}
	\caption{The $n$-hourglass-like band structures near the Fermi energy along the screw-invariant lines are shown for (a) BiPd, (b) ZnTe and (c) Tl$_3$PbBr$_5$. The full band structures and the crystal structures can be found in Ref.\cite{SM}. The spots and stars correspond to the same symbols in Fig.\ref{fig:graph}(a-c).}
	\label{fig:Bands}
\end{figure}

{\it Materials realization}.---To confirm our symmetry analysis and minimal model, we perform first-principles calculations of band structures on the non-magnetic materials BiPd, ZnTe under high pressure, and the high temperature phase of Tl$_3$PbBr$_5$, which belong to SG $P2_1$, $P3_1$ and $P4_1$, respectively. The details can be found in Ref.\cite{SM}, including the methods of calculations and the crystal structure and band structures of each material, with or without SOC. In Fig.\ref{fig:Bands} we show the band structures of the three materials along screw-invariant lines, where the $n$-hourglasses are seen for $n=2,3,4$. The Weyl points corresponding to those obtained by symmetry analysis are labeled by the same symbols as in Fig.\ref{fig:graph}. The band structures with Weyl points completely agree with the results predicted by symmetry analysis. Moreover, both the 1+1 and the 3+3 case of the distribution of Weyl points have been found in the band structure of ZnTe, indicated by the stars and spots in Fig.\ref{fig:Bands}(b), respectively.

The scenario how the $n$-hourglasses appear when SOC is turned on, shown by the minimal model above, is also confirmed by both band structure calculations and from compatibility relations, see Ref.\cite{SM}.

{\it Discussion and conclusion}.---Although the three representative materials we have discussed host $n$-hourglass Weyl fermions, they are not filling-enforced semimetals\cite{Watanabe2016}. To experimentally find the physical effects discussed, physical or chemical doping can be used to shift the Fermi energy towards the Weyl points, and searching for filling-enforced semimetals in the SGs specified may work better.

A future direction is to investigate the possible generalizations of the $n$-hourglasses to magnetic SGs, including the fermionic quasiparticle spectrum as well as the bosonic collective modes, magnons. As Weyl magnons have been found\cite{Li2016}, magnons with $n$-hourglass-like dispersion may appear in certain magnetic SGs. Metamaterials such as photonic crystals\cite{Lu2015} have proved to be a wonderful playground to realize topological phases, where we expect the $n$-hourglass Weyl points may also be engineered.

The idea of time crystals which exhibit discrete time translational symmetry has been conceived\cite{Wilczek2012} and objected\cite{Bruno2013,Watanabe2015}, and eventually a version different from the original one has been proved to work by both theories\cite{Else2016,Yao2017} and experiments\cite{Zhang2017}. The time crystals are classified according to the so-called space-time group symmetries\cite{Xu2018}. Here we mention that the nonsymmorphic space-time group symmetries, including the time-screw and time-glide, when combined with TRS, could have interesting features on the spectrum, though in different ways from the scenario shown in this work.

In conclusion, we have studied the crystalline solids that have a single screw axis and time-reversal symmetry. From symmetry analysis, we have shown that when a sizable SOC is present, the $n$-hourglass-like band structures are enforced along screw-invariant lines. Distinct physical properties including the quantum nonlinear Hall effect and topological density waves have been predicted for $n$-hourglass Weyl semimetals. A minimal model respecting all the symmtries has been constructed, using which we show how the $n$-hourglasses appear when the SOC is turned on. Our first-principles band structure calculations of the nonsymmorphic materials BiPd, ZnTe and Tl$_3$PbBr$_5$ have confirmed the symmetry analysis and minimal model. Our work provides a way to find symmetry-and-filling-enforced Weyl semimetals and to design Weyl metamaterials.

{\it Acknowledgement}.---L.W. thanks S.-K. Jian, H. Yao, Z. Yan and P. Ye for useful discussions. This work was supported by NKRDPC-2017YFA0206203, NSFC-11574404, NSFG-2015A030313176, National Supercomputer Center in Guangzhou and Leading Talent Program of Guangdong Special Projects.

\bibliography{Hourglass_Weyl_bib}

\end{document}


\title{Supplemental materials}

\maketitle

\section{The full band structures and methods of calculation}
Our first-principles calculations have been performed on BiPd, ZnTe under high pressure, and the high temperature phase of Tl$_3$PbBr$_5$, which possess a single $N$-fold screw axis with $N=2,3,4$, respectively, as well as TRS. The SOC is expected to be sizable in these materials.

\begin{figure}[ht]
	\includegraphics[width=8cm]{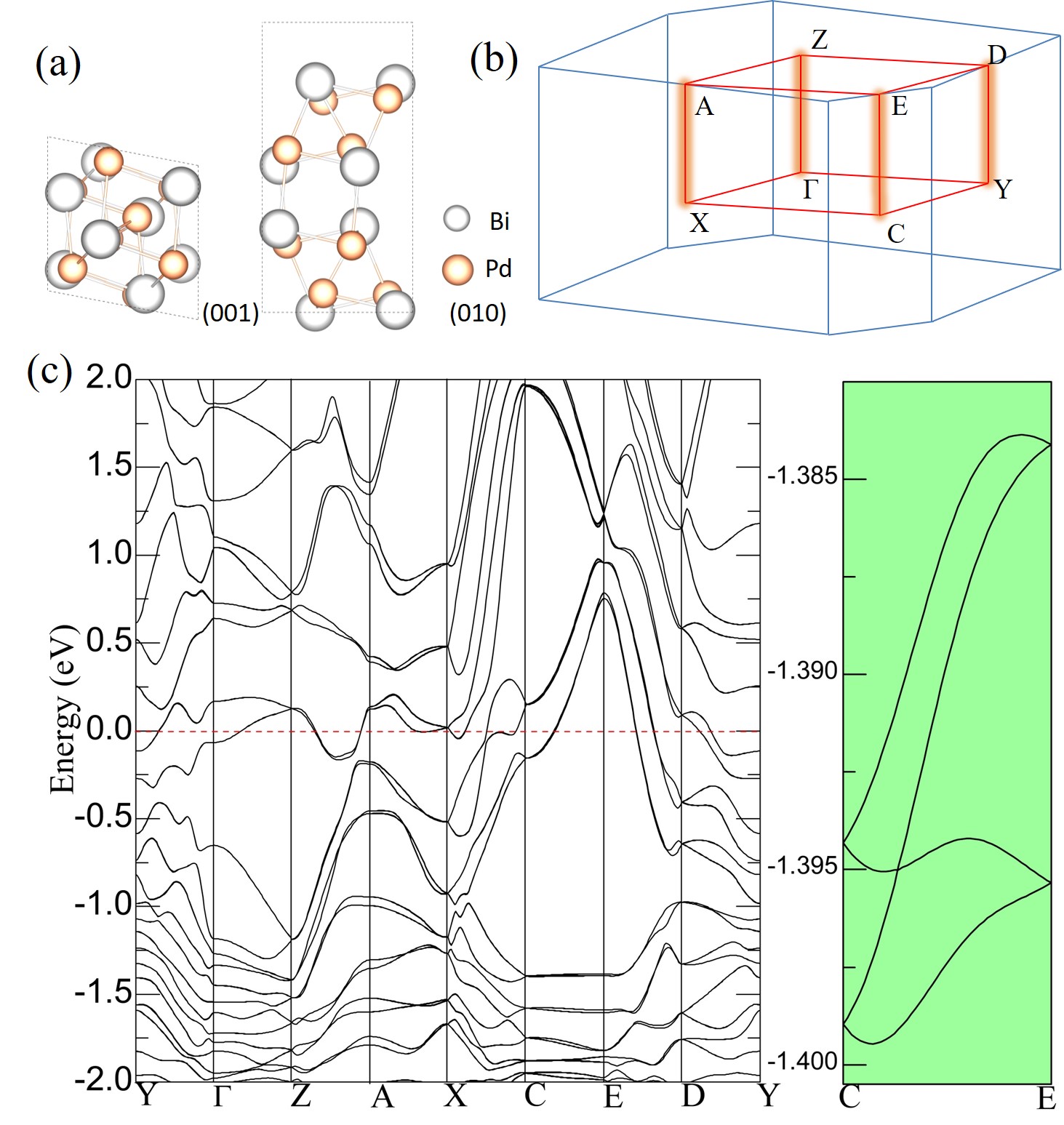}
	\caption{(a) Crystal structure of BiPd in SG $P2_1$. (b) The first BZ and the high symmetry points and lines. The screw-invariant lines are highlighted by glowing orange. (c) Band structure of BiPd (with SOC). A 2-hourglass along CE is enlarged and shown on the right.}
	\label{fig:BiPd}
\end{figure}

BiPd is a monoclinic crystal with 16 atoms in the unit cell. It becomes superconducting below 3.8 K\cite{PhysRevB.84.064518} and a Dirac cone surface state exists on the (010) surface\cite{BiPd1}. The band structure shows metallic behavior, with twelve bands crossing the Fermi energy. The SOC causes band splitting except at the eight TRIM where Kramers degeneracy appears. According to our symmetry analysis, the bands along the screw-invariant lines must show 2-hourglass-like dispersion. In Fig.\ref{fig:BiPd}, we show the full band structure of BiPd, and that along CE, a twofold screw-invariant line. It should be mentioned that all the bands, from the bottom valence bands to the top conduction bands show similar behavior -- every four bands form a 2-hourglass if the SOC is large enough.

\begin{figure}[ht]
		\includegraphics[width=8cm]{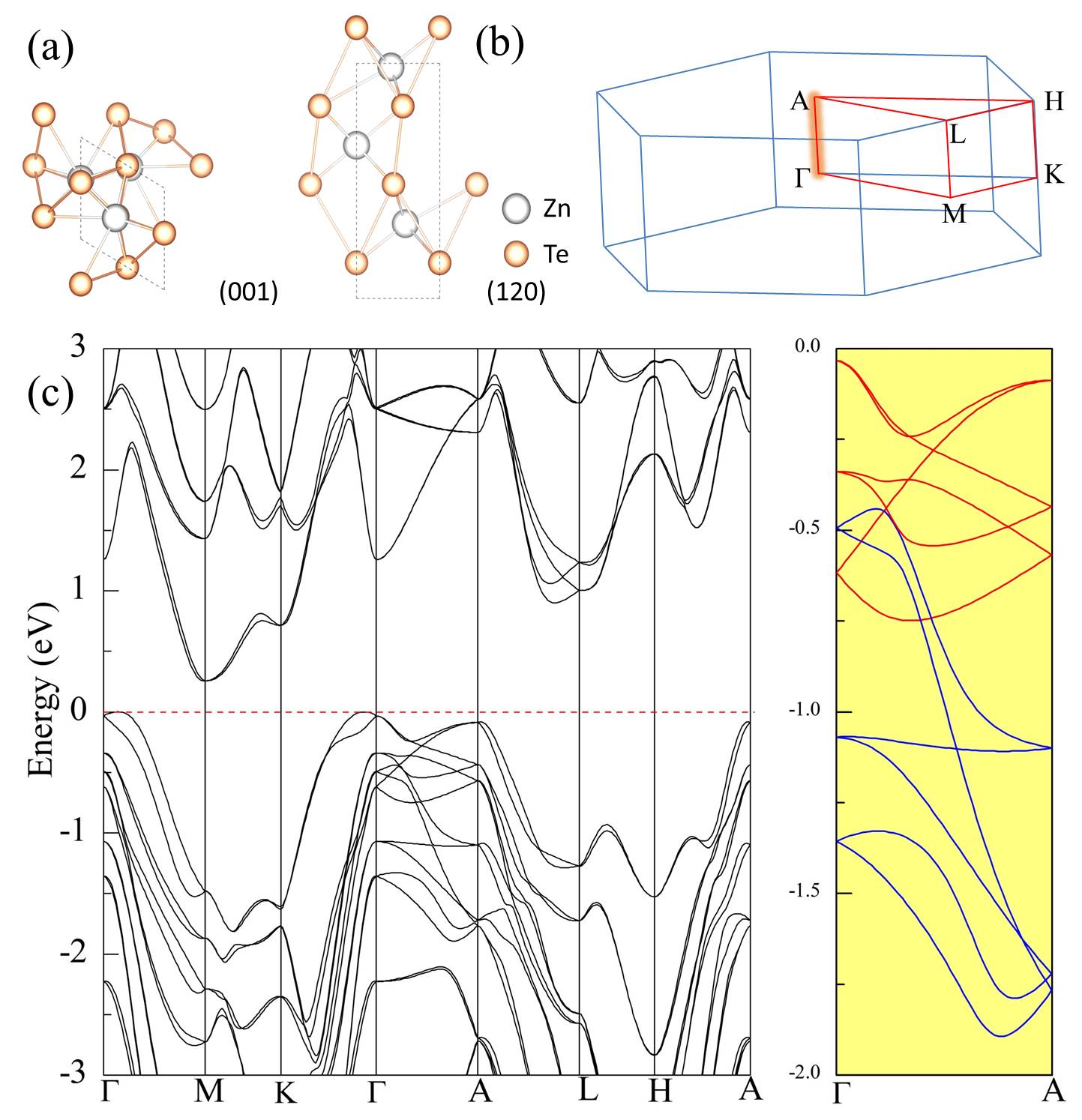}
		\caption{(a) Crystal structure of ZnTe in SG $P3_1$. (b) The first BZ and the high symmetry points and lines. The screw-invariant line ended with two TRIM is highlighted. (c) Band structure of ZnTe (with SOC). The valence bands near Fermi energy showing 3-hourglass band crossings are enlarged and shown on the right.}
		\label{fig:ZnTe}
\end{figure}

ZnTe is a \RNum{2}B-\RNum{6} semiconductor in zinc-blende (ZB) structure at ambient pressure, and turns into ZnTe-\RNum{2} phase at 9.6 GPa and further into ZnTe-\RNum{3} phase at 12.1 GPa\cite{ZnTe1}. The ZnTe-\RNum{3} phase is in the SG $Cmcm$ in the orthorhombic system, while controversy exists in identifying the ZnTe-\RNum{2} phase. Both cinnabar structure (SG $P3_121$)\cite{ZnTe3} and $P3_1$\cite{ZnTe2}(Fig.\ref{fig:ZnTe} (a)) have been claimed. Here we presume the $P3_1$ ZnTe exists and consider its band structures.

Similar to BiPd, 3-hourglass-like dispersions are observed for the bands along the TRIM-ended screw-invariant line $\Ga$A, as shown in Fig.\ref{fig:ZnTe}. Note that here the 3-hourglass is not standard due to the permutation of the vertices, thus more than two crossings occur. What is physically interesting here, as discussed in the main text, is that in half of the BZ, a minimal number of two Weyl points (the $1+1$ case) appears as the crossings of the eighth and the ninth band below the Fermi energy, and a minimal number of six Weyl points (the $3+3$ case) exists as the crossings of the second and the third band.

Tl$_3$PbBr$_5$ can achieve effective doping of rare-earth element and can be applied as non-linear optical devices\cite{Tl3PbBr5A}. Two phases are identified, the low-temperature orthorhombic phase (SG $P2_12_12_1$) and the high-temperature tetragonal phase (SG $P4_1$)\cite{Tl3PbBr5B}, with the transition temperature about 239 $^\circ$C. For the high-temperature phase there are 36 atoms per unit cell. 4-hourglass-like dispersions along the fourfold screw-invariant lines $\Ga$Z and MA, and 2-hourglass-like dispersions along the twofold screw-invariant line XR, are observed. A minimal number of two Weyl points (the $1+1$ case) might also exist for the fourth and the fifth band above the Fermi energy, one at $\Ga$Z and the other at MA.

Band structure calculations were performed by density functional theory using the Vienna Ab-initio Simulation Package (VASP)\cite{PhysRevB.54.11169}. The projector augmented-wave (PAW)\cite{PhysRevB.50.17953} pseudopotentials were used and the generalized gradient approximation of the PBE type functional\cite{PhysRevLett.77.3865} was used describing the exchange-correlation energy. A cutoff energy of 700 eV (500 eV for Tl$_3$PbBr$_5$) was used for the plane wave expansion to ensure the convergence of total energy to be less than 1 meV. The Monkhorst-Pack grid\cite{PhysRevB.13.5188} for BZ sampling was set to $7 \times 7 \times 7$, $15\times15\times9$, $7\times7\times 5$ for BiPd, ZnTe and Tl$_3$PbBr$_5$, respectively. The structures were fully optimized with experimental lattice constants used, until the forces were smaller than 0.01 eV/{\mbox{\AA}}.

\begin{figure}[ht]
	\includegraphics[width=8cm]{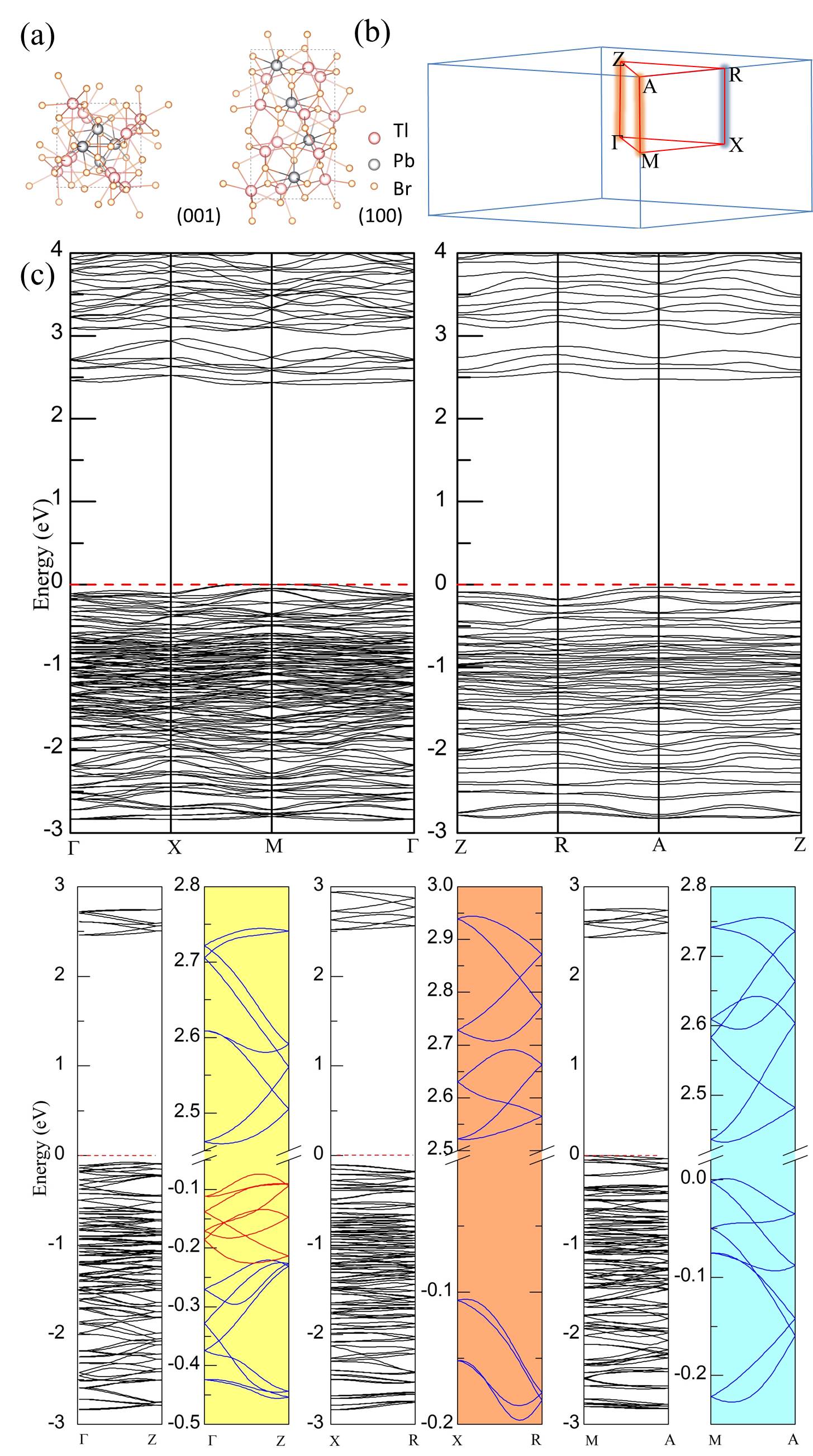}
	\caption{(a) Crystal structure of Tl$_3$PbBr$_5$ in SG $P4_1$. (b) The first BZ and high symmetry points and lines. The fourfold and twofold screw-invariant lines are highlighted by orange and blue, respectively. (c) Band structure of Tl$_3$PbBr$_5$ (with SOC). The conduction and valence bands near the Fermi energy showing $4$- and 2-hourglass band crossings are enlarged and shown on the right.}
	\label{fig:Tl3PbBr5}
\end{figure}

\section{Compatibility relations}
\label{sec:compat}
The {\it n}-hourglass-like band structures in the nonsymmorphic SGs $P2_1$,$P3_1$ and $P4_1$ can also be derived by considering the compatibility relatons\cite{PhysRevMaterials.2.074201} between irreducible representations at different high symmetry points and lines. We derive the compatibility relations for $P4_1$ and $P3_1$ below. We will see that the $n$-hourglasses are automatically generated as a rigorous result of the symmetry requirement.

Consider SG $P4_1$ first. For nonsymmorphic SGs, the SG representation $D(\{\alpha|\tau+t\})$ can be obtained from the corresponding point group representations $\Gamma(\{\alpha|0\})$ by $D(\{\alpha|\tau+t\})=e^{ik\tau}e^{ikt}\Gamma(\{\alpha|0\})$. For $\Ga (0,0,0)$, $\Lambda (0,0,\omega)$, and Z$(0,0,0.5\pi)$ in the BZ, the corresponding point groups -- or the group of the {\it k} vectors are the same, being $\{E,\{C_{2z}|\frac{1}{2}t_3\},\{C_{4z}|\frac{1}{4}t_3\},\{C_{4z}^{-1}|\frac{3}{4}t_3\}\}$. Since SOC is considered, the symmetry of bands are described by the double-valued representation\cite{Elcoro:ks5574} of the double SG. The corresponding character table for the double-valued representations of the group of wave vectors at $\Gamma$,$\Lambda$, and $Z$ are shown in Table.\ref{tab:P4_1}.

From the character table it is obvious that the representations at $\Gamma$ and Z are both doubly degenerate. This results in the double degeneracy of the bands at $\Gamma$ and Z, which are both TRIM. Along $\Lambda$, the representations are non-degenerate. By comparing the characters of each representation, it is easy to get the following compatibility relations between $\Gamma$ and $\Lambda$,
\begin{equation}
\begin{split}
\overline \Gamma_5 \overline \Gamma_7 \to \overline \Lambda_5 + \overline \Lambda_7, \\
\overline \Gamma_6 \overline \Gamma_8 \to \overline \Lambda_6 + \overline \Lambda_8,
\end{split}
\end{equation}
and the compatibility relations between Z and $\Lambda$,
\begin{equation}
\begin{split}
\overline Z_5 \overline Z_5 \to \overline \Lambda_5 + \overline \Lambda_5, \\
\overline Z_6 \overline Z_6 \to \overline \Lambda_6 + \overline \Lambda_6, \\
\overline Z_7 \overline Z_8 \to \overline \Lambda_7 + \overline \Lambda_8.
\end{split}
\end{equation}

\begin{table}
	\caption{The double-valued irreducible representations of SG $P4_1$, for the group of wave vectors at $\Gamma$, Z and $\Lambda$. The degeneracy due to TRS for each representation is added at the end of each row, with $a$ and $b$ meaning doubly degenerate and $c$ non-degenerate\cite{PhysRev.52.361}. Here $0<\omega<0.5$. For $\omega=0$, $\Lambda$ becomes $\Gamma$ and for $\omega=0.5$, $\Lambda$ becomes Z.}
	\begin{tabular}{c|ccccc}
		\hline
		Repres. & $E$ & $\{C_{2z}|\frac{1}{2}t_3\}$ & $\{C_{4z}|\frac{1}{4}t_3\}$ & $\{C_{4z}^{-1}|\frac{3}{4}t_3\}$ & T. I. \\
		\hline
		${\overline\Gamma_5}$ & $1$ & $-i$ & $e^{i\frac{3}{4}\pi}$ & $e^{-i\frac{3}{4}\pi}$ & $b$ \\
		${\overline\Gamma_6}$ & $1$ & $-i$ & $e^{-i\frac{1}{4}\pi}$ & $e^{i\frac{1}{4}\pi}$ & $b$ \\
		${\overline\Gamma_7}$ & $1$ & $i$ & $e^{-i\frac{3}{4}\pi}$ & $e^{i\frac{3}{4}\pi}$ & $b$ \\
		${\overline\Gamma_8}$ & $1$ & $i$ & $e^{i\frac{1}{4}\pi}$ & $e^{-i\frac{1}{4}\pi}$ & $b$ \\
		\hline
		${\overline Z_5}$ & $1$ & $1$ & $-1$ & $1$ & $a$ \\
		${\overline Z_6}$ & $1$ & $1$ & $1$ & $-1$ & $a$ \\
		${\overline Z_7}$ & $1$ & $-1$ & $-i$ & $-i$ & $b$ \\
		${\overline Z_8}$ & $1$ & $-1$ & $i$ & $i$ & $b$ \\
		\hline
		${\overline \Lambda_5}$ & $1$ & ${-ie^{i\pi \omega}}$ & ${e^{i\frac{3}{4}\pi}e^{i\frac{1}{2}\pi \omega}}$ & ${e^{-i\frac{3}{4}\pi}e^{i\frac{3}{2}\pi \omega}}$ & $c$ \\
		${\overline \Lambda_6}$ & $1$ & ${-ie^{i\pi \omega}}$ & ${e^{-i\frac{1}{4}\pi}e^{i\frac{1}{2}\pi \omega}}$ & ${e^{i\frac{1}{4}\pi}e^{i\frac{3}{2}\pi \omega}}$ & $c$ \\
		${\overline \Lambda_7}$ & $1$ & ${ie^{i\pi \omega}}$ & ${e^{-i\frac{3}{4}\pi}e^{i\frac{1}{2}\pi \omega}}$ & ${e^{i\frac{3}{4}\pi}e^{i\frac{3}{2}\pi \omega}}$ & $c$ \\
		${\overline \Lambda_8}$ & $1$ & ${ie^{i\pi \omega}}$ & ${e^{i\frac{1}{4}\pi}e^{i\frac{1}{2}\pi \omega}}$ & ${e^{-i\frac{1}{4}\pi}e^{i\frac{3}{2}\pi \omega}}$ & $c$ \\
		\hline		
	\end{tabular}
	\label{tab:P4_1}
\end{table}

\begin{figure}[ht]
	\includegraphics[width=8cm]{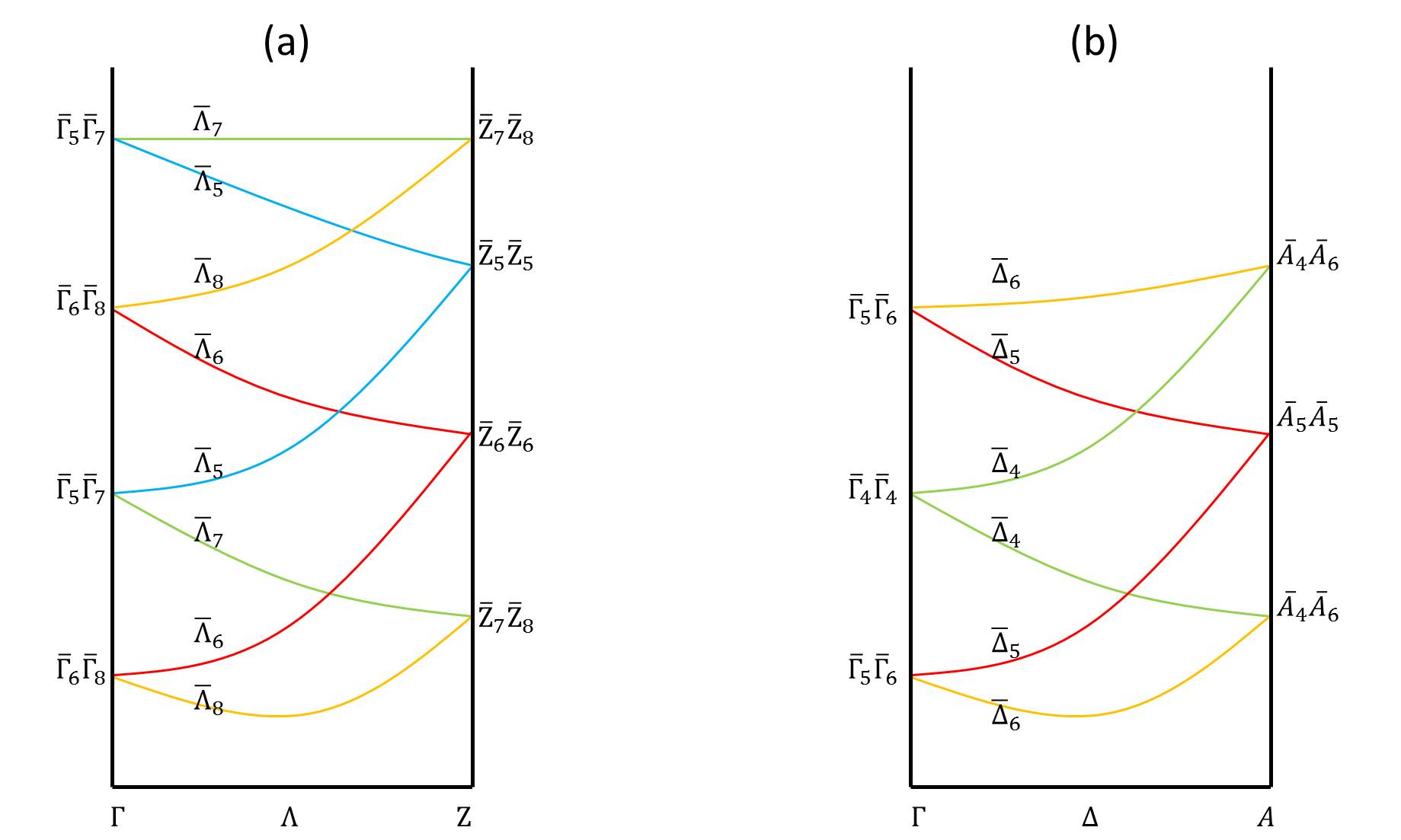}
	\caption{(a) The compatibility relations between bands at $\Gamma$, $\Lambda$ and Z of crystals with $P4_1$ SG symmetry. (b) The compatibility relations between bands at $\Gamma$, $\Delta$ and A of crystals with $P3_1$ SG symmetry.}
	\label{fig:compat}
\end{figure}
The compatibility relations can be schematically shown in Fig.\ref{fig:compat}(a), which result in the band structure similar to that obtained in the main text by symmetry analysis. We see that each representation has to appear twice to make the set of bands closed, i.e. separated from another set. In Fig.\ref{fig:compat}(a), they are arranged in such a way to result in a standard 4-hourglass band structure with three crossings along $\Lambda$. By tuning the sequence of the representations at $\Gamma$ or Z we can get a variant of the standard 4-hourglass with more crossings.

Similarly, the double-valued irreducible representations of SG $P3_1$ for the group of wave vectors at $\Gamma$, $\Delta$ and A are shown in Table.\ref{tab:P3_1}, from which we have the compatibility relations between $\Gamma$ and $\Delta$,
\begin{equation}
\begin{split}
\overline \Gamma_4 \overline \Gamma_4 \to \overline \Delta_4 + \overline \Delta_4, \\
\overline \Gamma_5 \overline \Gamma_6 \to \overline \Delta_5 + \overline \Delta_6,
\end{split}
\end{equation}
and the compatibility relation between A and $\Delta$
\begin{equation}
\begin{split}
\overline A_4 \overline A_6 \to \overline \Delta_4 + \overline \Delta_6, \\
\overline A_5 \overline A_5 \to \overline \Delta_5 + \overline \Delta_5.
\end{split}
\end{equation}
The compatibility relations are shown schematically in Fig.\ref{fig:compat}(b), with minimally two crossings along $\Delta$.

\begin{table}
	\caption{Double-valued irreducible representations of SG $P3_1$ for the group of wave vectors at $\Gamma (0,0,0)$, A$(0,0,0.5)$ and $\Delta(0,0,\omega)$ with $0<\omega<0.5$. The degeneracy due to TRS for each representation is added at the end of each row, with $a$ and $b$ meaning doubly degenerate and $c$ non-degenerate.}
	\begin{tabular}{c|cccc}
		\hline
		Repres. & $E$ & $\{C_{3z}|\frac{1}{3}t_3\}$ & $\{C_{3z}^{-1}|\frac{2}{3}t_3\}$ & T. I. \\
		\hline
		$\overline \Gamma_4$ & $1$ & $-1$ & $-1$ & $a$ \\
		$\overline \Gamma_5$ & $1$ & $e^{-i\frac{1}{3}\pi}$ & $e^{i\frac{1}{3}\pi}$ & $b$ \\
		$\overline \Gamma_6$ & $1$ & $e^{i\frac{1}{3}\pi}$ & $e^{-i\frac{1}{3}\pi}$ & $b$ \\
		\hline
		$\overline A_4$ & $1$ & $e^{-i\frac{2}{3}\pi}$ & $e^{-i\frac{1}{3}\pi}$ & $b$ \\
		$\overline A_5$ & $1$ & $1$ & $-1$ & $a$ \\
		$\overline A_6$ & $1$ & $e^{i\frac{2}{3}\pi}$ & $e^{i\frac{1}{3}\pi}$ & $b$ \\
		\hline
		$\overline \Delta_4$ & $1$ & $-e^{i\frac{2}{3}\pi \omega}$ & $-e^{i\frac{4}{3}\pi \omega}$ & $c$ \\
		$\overline \Delta_5$ & $1$ & $e^{-i\frac{1}{3}\pi}e^{i\frac{2}{3}\pi \omega}$ & $e^{i\frac{1}{3}\pi}e^{i\frac{4}{3}\pi \omega}$ & $c$ \\
		$\overline \Delta_6$ & $1$ & $e^{i\frac{1}{3}\pi}e^{i\frac{2}{3}\pi \omega}$ & $e^{-i\frac{1}{3}\pi}e^{i\frac{4}{3}\pi \omega}$ & $c$ \\
		\hline
	\end{tabular}
	\label{tab:P3_1}
\end{table}

\section{Band structures without SOC}
The $n$-hourglass-like band structures appear in the presence of SOC. In the main text, we have pointed out that when SOC is ignored, the $n$-hourglass-like structures turn into V-like, N-like and W-like structures for $n=2,3,4$, respectively, using our minimal model. Here we give the band structures without SOC and show that the V-, N-, and W-like band structures can also be understood from the compatibility relations between the single-valued irreducible representations of the group of wave vectors.

\begin{table}[ht]
	\caption{The single-valued irreducible representations of SG $P4_1$, for the group of wave vectors at $\Gamma$, Z and $\Lambda$. The degeneracy due to TRS for each representation is added at the end of each row, with $a$ meaning non-degenerate and $b$,$c$ doubly degenerate. Here $0<\omega<0.5$. For $\omega=0$, $\Lambda$ becomes $\Gamma$ and for $\omega=0.5$, $\Lambda$ becomes Z.}
	\begin{tabular}{c|ccccc}
		\hline
		Repres. & $E$ & $\{C_{2z}|\frac{1}{2}t_3\}$ & $\{C_{4z}|\frac{1}{4}t_3\}$ & $\{C_{4z}^{-1}|\frac{3}{4}t_3\}$ & T. I. \\
		\hline
		${\Gamma_1}$ & $1$ & $1$ & $1$ & $1$ & $a$ \\
		${\Gamma_2}$ & $1$ & $1$ & $-1$ & $-1$ & $a$ \\
		${\Gamma_3}$ & $1$ & $-1$ & $i$ & $-i$ & $b$ \\
		${\Gamma_4}$ & $1$ & $-1$ & $-i$ & $i$ & $b$ \\
		\hline
		${Z_1}$ & $1$ & $i$ & $e^{i\frac{1}{4}\pi}$ & $e^{i\frac{3}{4}\pi}$ & $b$ \\
		${Z_2}$ & $1$ & $i$ & $e^{-i\frac{3}{4}\pi}$ & $e^{-i\frac{1}{4}\pi}$ & $b$ \\
		${Z_3}$ & $1$ & $-i$ & $e^{i\frac{3}{4}\pi}$ & $e^{i\frac{1}{4}\pi}$ & $b$ \\
		${Z_4}$ & $1$ & $-i$ & $e^{-i\frac{1}{4}\pi}$ & $e^{-i\frac{3}{4}\pi}$ & $b$ \\
		\hline
		${\Lambda_1}$ & $1$ & ${e^{i\pi \omega}}$ & ${e^{i\frac{1}{2}\pi \omega}}$ & ${e^{i\frac{3}{2}\pi \omega}}$ & $a$ \\
		${\Lambda_2}$ & $1$ & ${e^{i\pi \omega}}$ & ${-e^{i\frac{1}{2}\pi \omega}}$ & ${-e^{i\frac{3}{2}\pi \omega}}$ & $a$ \\
		${\Lambda_3}$ & $1$ & ${-e^{i\pi \omega}}$ & ${ie^{i\frac{1}{2}\pi \omega}}$ & ${-ie^{i\frac{3}{2}\pi \omega}}$ & $a$ \\
		${\Lambda_4}$ & $1$ & ${-e^{i\pi \omega}}$ & ${-ie^{i\frac{1}{2}\pi \omega}}$ & ${ie^{i\frac{3}{2}\pi \omega}}$ & $a$ \\
		\hline		
	\end{tabular}
	\label{tab:P4_1_single_valued}
\end{table}
\begin{figure}
	\includegraphics[width=8cm]{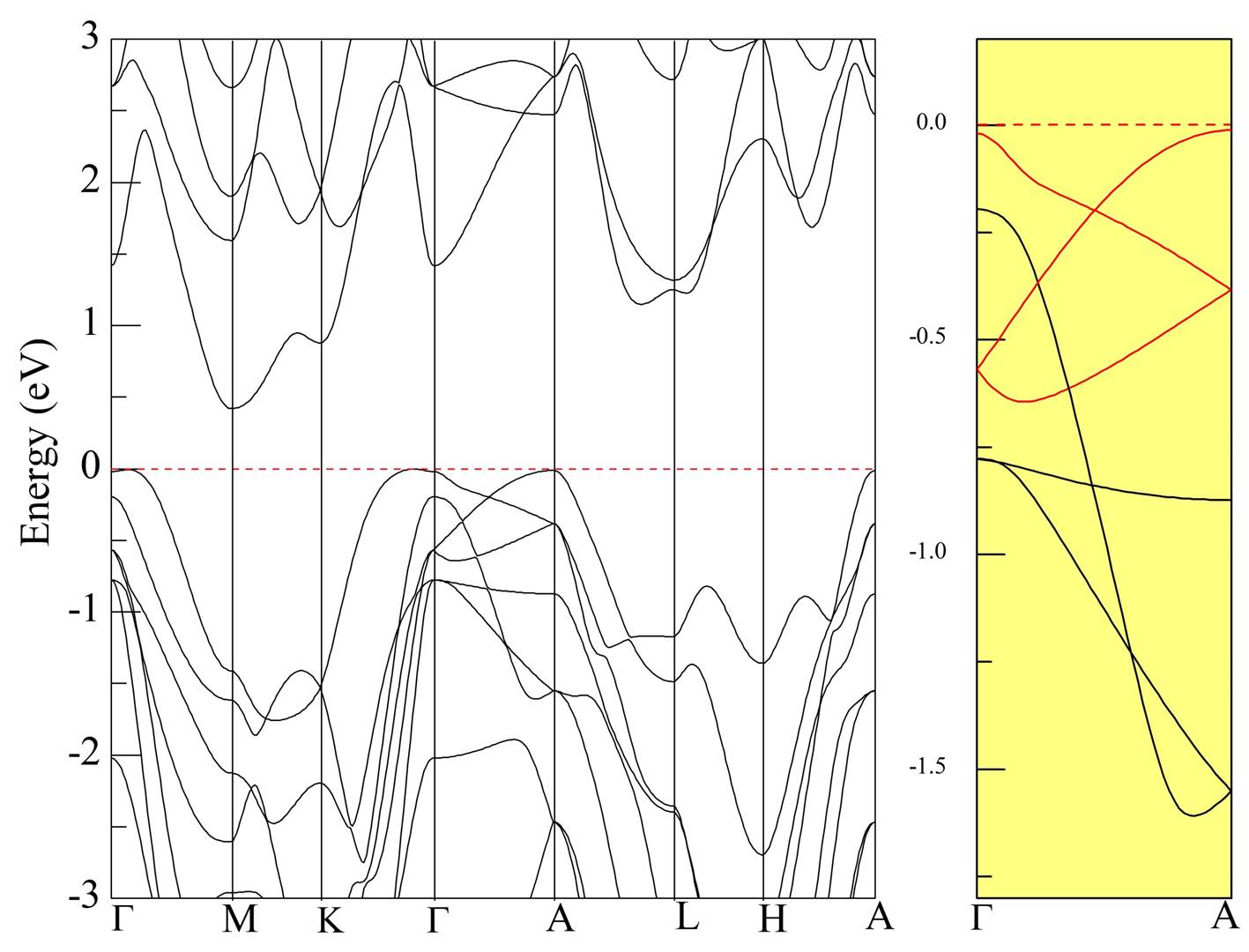}
	\caption{Band structure of ZnTe without SOC. The valence bands near Fermi energy showing N-like dispersions are enlarged and shown on the right.}
	\label{fig:ZnTe_woSOC}
\end{figure}
\begin{figure}
	\includegraphics[width=8cm]{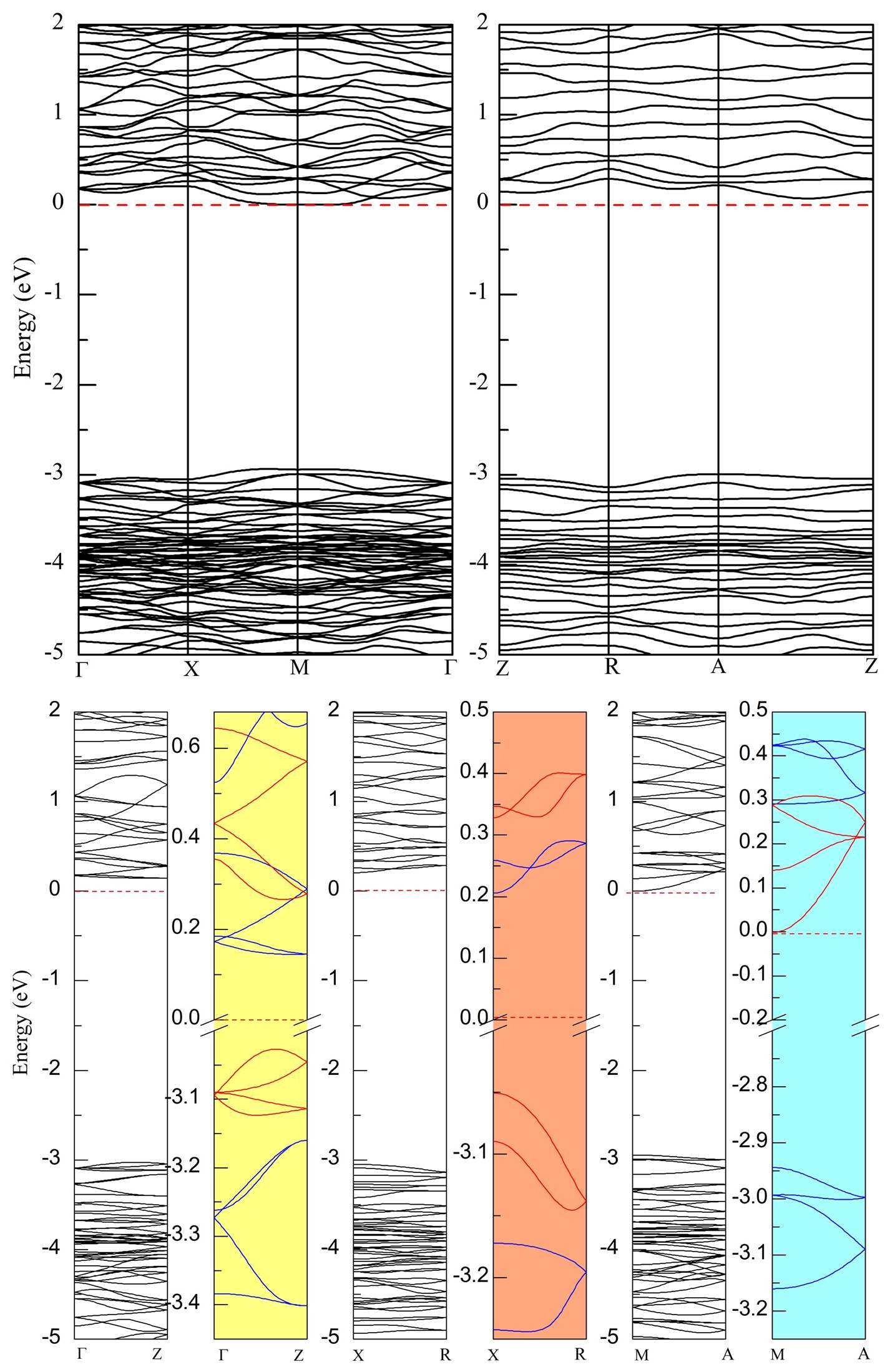}
	\caption{Band structure of Tl$_3$PbBr$_5$ without SOC. The conduction and valence bands near Fermi energy showing W-like dispersions are enlarged and shown on the right of each screw-invariant path.}
	\label{fig:Tl3PbBr5_woSOC}
\end{figure}

\begin{figure}
	\includegraphics[width=8cm]{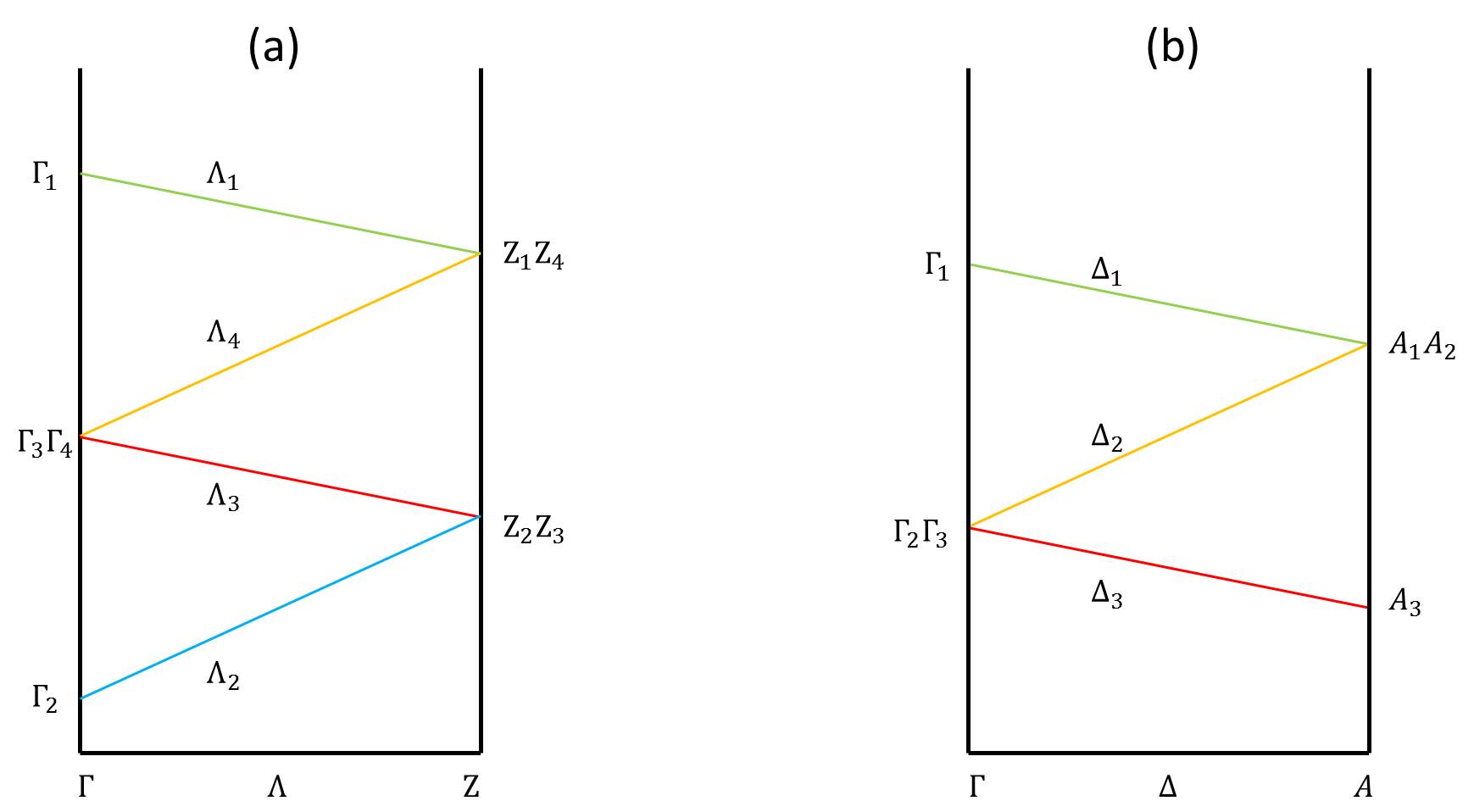}
	\caption{(a) The compatibility relations between bands at $\Gamma$, $\Lambda$ and $Z$ of crystals with $P4_1$ SG, ignoring SOC. (b) Compatibility relation for bands at $\Gamma$, $\Delta$ and $A$ of crystals with $P3_1$ SG, also ignoring SOC.}
	\label{fig:compat_woSOC}
\end{figure}

For SG $P4_1$, the groups of wave vectors at $\Gamma$, Z and $\Lambda$ are the same with that in Sec.\ref{sec:compat}. The only difference is that now we are concerned with the single-valued irreducible representations\cite{Tinkham}, which describe the symmetry of bands when the Hamiltonian contains no spin-dependent terms. The results are shown in Tab.\ref{tab:P4_1_single_valued}, from which we can derive the compatibility relations between irreducible representations of the group of wave vectors at $\Gamma$ and $\Lambda$,

\begin{equation}
\begin{split}
\Gamma_1 \to \Lambda_1, \\
\Gamma_2 \to \Lambda_2, \\
\Gamma_3 \Gamma_4 \to \Lambda_3 + \Lambda_4,
\end{split}
\end{equation}
and the compatibility relations between the irreducible representation at Z and $\Lambda$,
\begin{equation}
\begin{split}
Z_1 Z_4 \to \Lambda_1 + \Lambda_4, \\
Z_2 Z_3 \to \Lambda_2 + \Lambda_3.
\end{split}
\end{equation}
From the compatibility relations, the W-like band structure along $\Ga$Z can be deduced and is shown schematically in Fig.\ref{fig:compat_woSOC}(a).

\begin{table}
	\caption{Single-valued irreducible representations of SG $P3_1$ for the group of wave vectors at $\Gamma (0,0,0)$, A$(0,0,0.5)$ and $\Delta(0,0,\omega)$, with $0<\omega<0.5$. The degeneracy due to TRS for each representation is added at the end of each row, with $a$ meaning non-degenerate and $b$ and $c$ doubly degenerate.}
	\begin{tabular}{c|cccc}
		\hline
		Repres. & $E$ & $\{C_{3z}|\frac{1}{3}t_3\}$ & $\{C_{3z}^{-1}|\frac{2}{3}t_3\}$ & T. I. \\
		\hline
		$\Gamma_1$ & $1$ & $1$ & $1$ & $a$ \\
		$\Gamma_2$ & $1$ & $e^{i\frac{2}{3}\pi}$ & $e^{-i\frac{2}{3}\pi}$ & $b$ \\
		$\Gamma_3$ & $1$ & $e^{-i\frac{2}{3}\pi}$ & $e^{i\frac{2}{3}\pi}$ & $b$ \\
		\hline
		$A_1$ & $1$ & $e^{i\frac{2}{3}\pi}$ & $e^{-i\frac{2}{3}\pi}$ & $b$ \\
		$A_2$ & $1$ & $e^{-i\frac{2}{3}\pi}$ & $e^{i\frac{2}{3}\pi}$ & $b$ \\
		$A_3$ & $1$ & $1$ & $1$ & $a$ \\
		\hline
		$\Delta_1$ & $1$ & $e^{i\frac{2}{3}\pi \omega}$ & $e^{i\frac{4}{3}\pi \omega}$ & $a$ \\
		$\Delta_2$ & $1$ & $e^{i\frac{2}{3}\pi}e^{i\frac{2}{3}\pi \omega}$ & $e^{-i\frac{2}{3}\pi}e^{i\frac{4}{3}\pi \omega}$ & $a$ \\
		$\Delta_3$ & $1$ & $e^{-i\frac{2}{3}\pi}e^{i\frac{2}{3}\pi \omega}$ & $e^{i\frac{2}{3}\pi}e^{i\frac{4}{3}\pi \omega}$ & $a$ \\
		\hline
	\end{tabular}
    \label{tab:P3_1_single_valued}
\end{table}

In a similar way we can obtain  the single-valued irreducible representation of crystals with $P3_1$ SG symmetry, as shown in Table.\ref{tab:P3_1_single_valued}, from which the compatibility relations between $\Gamma$ and $\Delta$,
\begin{equation}
\begin{split}
\Gamma_1 \to \Delta_1, \\
\Gamma_2 \Gamma_3 \to \Delta_2 + \Delta_3,
\end{split}
\end{equation}
and the compatibility relations between A and $\Delta$,
\begin{equation}
\begin{split}
	A_1 A_2 \to \Delta_1 + \Delta_2, \\
	A_3 \to \Delta_3,
\end{split}
\end{equation}
are obtained. The N-like band structures along $\Gamma$A can be deduced, as shown in Fig.\ref{fig:compat_woSOC}(b).

The band structures of ZnTe and Tl$_3$PbBr$_5$ without SOC are shown in Fig.\ref{fig:ZnTe_woSOC} and Fig.\ref{fig:Tl3PbBr5_woSOC}, respectively. For ZnTe, along the screw-invariant line ($\Gamma$A), the band structure shows N-like dispersion. The shape can vary due to permutations of the vertices at $\Gamma$ and $A$ as well as curving. Similarly, for Tl$_3$PbBr$_5$, along $\Gamma$Z and MA, the band structure show W-like dispersions. Along XR, which is a twofold screw-invariant line, the bands show V-like dispersion. In the whole plane ZRA the bands are doubly degenerate, due to the presence of a twofold screw axis\cite{Wang2017PRB}.

\bibliography{Hourglass_Weyl_bib}